\documentclass[12pt]{iopart}

\usepackage{units}
\usepackage{iopams}  
\newcommand{\x}{\mathbf{x}}
\newcommand{\vel}{\mathbf{v}}
\newcommand{\kvec}{\mathbf{k}}

\usepackage{graphicx}
\usepackage{caption}

\usepackage{fixme}
\fxsetup{
    status=draft,
    author=,
    layout=inline,
    theme=color
}

\begin{document}

\title[]{A Spectral Canonical Electrostatic Algorithm}

\author{Stephen D. Webb$^1$}

\address{$^1$RadiaSoft, LLC, 1348 Redwood Ave., Boulder CO 80304}

\ead{swebb@radiasoft.net}

\begin{abstract}
Studying single-particle dynamics over many periods of oscillations is a well-understood problem solved using symplectic integration. Such integration schemes derive their update sequence from an approximate Hamiltonian, guaranteeing that the geometric structure of the underlying problem is preserved. Simulating a self-consistent system over many oscillations can introduce numerical artifacts such as grid heating. This unphysical heating stems from using non-symplectic methods on Hamiltonian systems. With this guidance, we derive an electrostatic algorithm using a discrete form of Hamilton's Principle. The resulting algorithm, a gridless spectral electrostatic macroparticle model, does not exhibit the unphysical heating typical of most particle-in-cell methods. We present results of this using a two-body problem as an example of the algorithm's energy- and momentum-conserving properties.
\end{abstract}

\vspace{2pc}

\maketitle

\section{Introduction}

The numerical simulation of plasmas using macroparticle models\footnote{
We distinguish between a macroparticle model -- which represents large numbers of point particles with a single particle of finite spatial extent -- and a particle-in-cell model -- which places those finite spatial extent particles into a spatial grid and carries out charge deposition by interpolating to the grid. Thus, particle-in-cell models are a subset of macroparticle models. In spectral algorithms, both options are viable, although in this paper we consider only gridless methods.
}~\cite{hockney_eastwood:89, birdsall_langdon:85} is a common tool for applications ranging from plasma processing to particle accelerators, fusion to laser-plasma interactions. The traditional approach to such algorithms starts with discretizing the Lorentz force law in time, assuming the fields are given everywhere in space. The Poisson or Maxwell equations are then discretized in space and time, and updated synchronous to when the Lorentz force integrator requires them. The fields used for the Lorentz force law are interpolated from the discrete spatial points at the proper time. Depositing the source terms is then chosen to carefully avoid self-forces on a single particle and conserve the local charge density. The update sequence is also frequently chosen to conserve momentum, energy, or both.

This process is prone to unphysical heating due to various instabilities having to do with the grid, the time discretization, or both. They also impose conservation laws which do not arise naturally from the discretization. Essentially, the problem is that when discretizing the equations of motion there are too many degrees of freedom left to the algorithm designer which may lead to unphysical behavior. 

This can be remedied by deriving the algorithms from a Hamiltonian least-action principle. Such multisymplectic algorithms, described most recently by Shadwick, Stamm, and Evstatiev~\cite{shadwick:14a, shadwick:14b}, respect the fact that the dynamics of both particles and fields~\cite{marsdenPatrickShkoller:98,marsdenPekarskyShkollerWest:01} have a symplectic geometry. The benefits of symplectic integration for single-particle mechanics are well-known in the accelerator physics community (see, for example, \cite{forest:06} for an historical overview of the subject and its applications). The idea of multisymplectic particle-in-cell algorithms has generated considerable interest, and a number of one-dimensional examples on a discrete spatial grid have recently been published~\cite{kraus:13,qin:15a,squire_etal:12,xiao_etal:15}

The chief benefit of multisymplectic algorithms is that their solutions must be a solution for some Lagrangian which approximates the continuous, ``real world'' Lagrangian. The numerical solutions are exact solutions to an approximate Lagrangian. This contributes to long-time stability over many oscillations of the system. In the case of plasmas, such algorithms promise to make simulations extremely reliable over hundreds or even thousands of plasma periods.

By making the discretizing approximations in the action integral, and then deriving the resulting equations of motion, we remove many of the undetermined components that can appear when approximations are made to the equations of motion directly. For example, charge deposition and force interpolation arise from a common term in the Lagrangian, and thus any approximations made to the particles and fields will relate these two constructively. The textbook method is to insist on a deposition/interpolation scheme which does not produce self-forces on the particles. This may or may not be consistent with the underlying physics which produces the equations of motion.

What is important to understand here is that the familiar physical principles -- conservation of momentum, conservation of energy, conservation of angular momentum, and on -- are encoded not in the equations of motion but in the symmetries of the action. They appear in the equations of motion, but arise because of symmetries under translation in space and time, rotations, and on, per Noether's Theorem. If we want our algorithms to contain these conservation laws consistently, then our algorithms should come from some action principle which has the appropriate symmetries.

In this paper we introduce an electrostatic canonical gridless macroparticle algorithm using a spectral decomposition of the electrostatic fields. We express the Lagrangian in terms of the electrostatic potential and particle phase space density (Section~\ref{lowLagrangian}), then discretize the Lagrangian into Fourier modes for the fields and macro-particles for the phase space density (Section~\ref{macroparticles}). In Section~\ref{algorithm}, the resulting action integral is discretized into a second order Riemann sum, and we use the discrete Euler-Lagrange equations to arrive at a second order in time spectral algorithm. The linear numerical dispersion relation is derived in Section~\ref{numerical_dispersion}. We demonstrate the momentum and energy conservation properties for a two-dimensional two-body problem in Section~\ref{twobody} and for a thermal plasma in Section~\ref{thermal_plasma}. Finally, we discuss some optimization strategies for implementation in Section~\ref{implementation}.

\section{The electrostatic Lagrangian} \label{lowLagrangian}

Single particle symplectic algorithms can be derived either from a sequence of canonical transformations~\cite{ruth:83}, splitting methods or so-called ``kick codes''~\cite{schachinger_talman:87}, or from a discrete action minimization~\cite[and references therein]{marsden_west:01}. Regardless of the formalism used to derive the symplectic algorithm, the numerical solutions are an exact solution to the equations of motion derived from an approximate action. This is a much tighter constraint than schemes which start with the equations of motion, where any arbitrary discretization scheme likely fails to respect the geometric properties required by a system described by a Hamiltonian least-action principle. This is what gives them their long-term stability.

In this paper, we consider the Lagrangian treatment of the least-action principle. This contains the same geometric structures as a Hamiltonian approach, but will be more convenient for future work with electromagnetics, where the canonical momentum's relation to the vector potential makes explicit integration schemes much more difficult.

We begin with the electrostatic Lagrangian for a system of particles and fields, due to Low\footnote{
We use C.G.S. units throughout this paper.
}~\cite{low:58}
\begin{eqnarray} \label{lowlagrangian}
\nonumber\mathcal{L} =& \int d\x_0 d \vel_0 \left [ \frac{1}{2} m \left ( \frac{\partial \x}{\partial t}(\x_0, \vel_0) \right )^2 - e \varphi(\x)\right ] f(\x_0, \vel_0) + \dots \\
&\dots + \frac{1}{8 \pi} \int d\x \nabla \varphi \cdot \nabla \varphi .
\end{eqnarray}
An auxiliary requirement is that the phase space volume is conserved, \emph{i.e.} $f(\x_0, \vel_0, 0) = f(\x, \vel, t)$, which is a statement of the Vlasov equation. The continuum equations of motion come from minimizing the action $\mathcal{S} = \int dt \mathcal{L}$ with respect to $\x$ and $\varphi$, and the resulting equations are the familiar Newton's Second Law for the particle characteristics, and the Poisson equation:
\numparts
\begin{eqnarray}
\frac{\partial}{\partial t} \frac{\delta \mathcal{L}}{\delta \x'} - \frac{\delta \mathcal{L}}{\delta \x} = 0 \rightarrow m \frac{\partial^2 \x}{\partial t^2} + e\nabla \varphi(\x) = 0 \\
\frac{\partial}{\partial_\mu} \frac{\delta \mathcal{L}}{\delta \varphi_\mu} - \frac{\delta \mathcal{L}}{\delta \varphi} = 0 \rightarrow\nabla^2 \varphi = 4 \pi e \int d \vel f(\x, \vel, t)
\end{eqnarray}
\endnumparts

To derive our discrete equations of motion from an action, we must first decompose $\mathcal{L}$ into a discrete set of coefficients the basis functions of our phase space density and scalar potential, and then approximate $\mathcal{S}$ as some discrete approximation of the action integral. The former is a mode decomposition, while the later is an approximation of the action integral using Riemann sums. Once this is done, we will use the discrete Euler-Lagrange equations from~\cite{marsden_west:01} to get the discrete equations of motion.

\section{Spatial discretization of the Lagrangian}\label{macroparticles}

Two fields must be discretized -- the phase space density and the scalar potential. Discretizing $\varphi$ will specify the Poisson solving algorithm, while discretizing the phase space density will give us a definition of macroparticles.

We discretize the scalar potential in a discrete Fourier-type basis, allowing for shape functions in Fourier space:
\begin{equation}
\varphi(\x) = \frac{1}{\sqrt{2 \pi}^D}\int d\kvec \sum_{\sigma} e^{i \kvec \cdot \x} \Psi(\kvec - \kvec_\sigma) \tilde{\varphi}_\sigma
\end{equation}
where $\sigma$ is a collective index for the discrete modes and $\Psi$ is a generic shape function for the Fourier modes.

\begin{figure}
\centering
\includegraphics[scale=0.5]{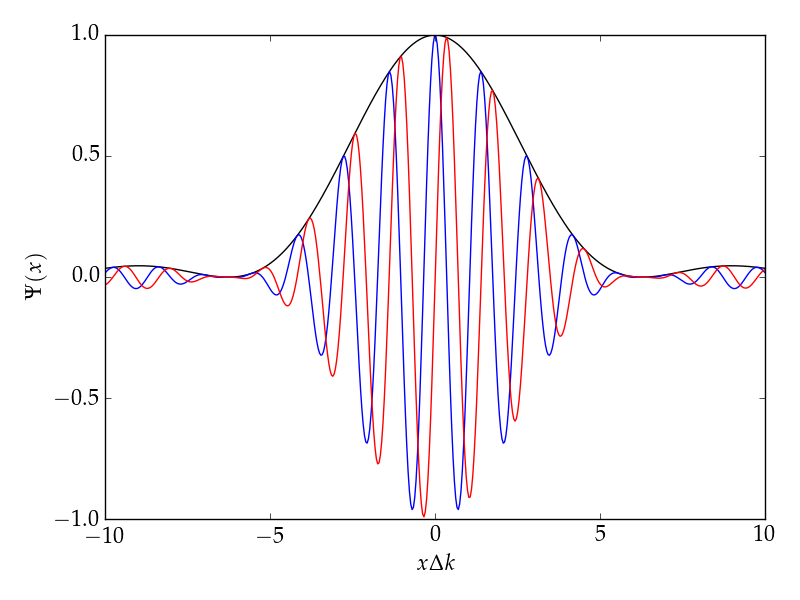}
\caption{The real and imaginary parts in the $x$ space for a tent shape function $\Psi(k - k_\sigma)$. The envelope drops off as $(x ~\Delta k)^{-2}$ for the tent function.}
\label{tentfunction}
\end{figure}

The choice of spectral shape function can be made to minimize ringing. For example, a delta function will yield the unattenuated $\cos (k_\sigma x)$ oscillation. By using tent functions separated by $\Delta k$, on the other hand, it is possible to accurately represent the Fourier transform of the fields in a piecewise linear fashion. This puts an envelope on the oscillations, and damps out ringing. Other options are also possible in this formalism, although we must consider their specific properties, as we do in \S~\ref{algorithm} below. In this way, the delta function option may prove the fastest, although it has some issues with ringing. An example of the tent function shape in real space is given in figure (\ref{tentfunction}).

To discretize the phase space density, we take shape functions in real space for the positions, and use delta functions in velocity:
\begin{equation}
f(\x, \vel, t) = \sum_j w_j \Lambda(\x - \x_j(t)) \delta(\vel - \vel_j(t))
\end{equation}
where $\x_j$ and $\vel_j$ are the co\"{o}rdinates and velocities for the $j^{th}$ macroparticle, which has a weight $w_j$. Here, we use $\vel$ as shorthand for $\partial_t \x$. We can understand macroparticles in this context as a finite element Lagrangian picture of the flow of the phase space fluid.

Inserting these expressions into the action, eqn. (\ref{lowlagrangian}), yields a discrete Lagrangian whose independent dynamical variables are the Fourier mode coefficients of the scalar potential and the co\"ordinates and velocities of the macroparticles:
\begin{eqnarray}
\nonumber \mathbf{L}_D =& \sum_j w_j \left \{ \frac{1}{2} m \left ( \frac{\partial \x_j}{\partial t} \right )^2 - e\sum_\sigma \int d\kvec \tilde{\Lambda}^*(\kvec)  e^{i \kvec \cdot \x_j} \Psi(\kvec - \kvec_\sigma) \tilde{\varphi}_\sigma \right \} \\
&+ \frac{1}{8 \pi} \sum_\sigma \sum_{\sigma'} \tilde{\varphi}_\sigma \tilde{\varphi}_{\sigma'} \int d \kvec ~ \kvec \cdot \kvec \Psi(\kvec - \kvec_{\sigma}) \Psi(-\kvec - \kvec_{\sigma'}).
\end{eqnarray}
We have used the orthogonality of the Fourier modes when integrating over $\x$ in the scalar potential energy term, and the particle shape function Fourier transform, $\tilde{\Lambda}(\kvec)$, is defined the same as the Fourier transform for $\varphi$ above. Specifically, this requires that
\begin{equation}
\Lambda(\x) = \frac{1}{\sqrt{2 \pi}^D}\int d\kvec e^{i \kvec \cdot \x} \tilde{\Lambda}(\kvec) \rightarrow \tilde{\Lambda}(\kvec) = \frac{1}{\sqrt{2 \pi}^D} \int d\x e^{-i \kvec \cdot \x} \Lambda(\x)
\end{equation}
from the Dirac delta function identity $\delta(\x) = (2 \pi)^D \int d\kvec e^{i \kvec \cdot \x}$. For convenience we will define the $k$-squared matrix as
\begin{equation}
\mathcal{K}_{\sigma \sigma'} \equiv \int d \kvec ~ \kvec \cdot \kvec \Psi(\kvec - \kvec_\sigma) \Psi(-\kvec - \kvec_{\sigma'}).
\end{equation}

If we consider the simple discrete Fourier transform, with $\Psi(\kvec - \kvec_\sigma) = \delta (\kvec - \kvec_\sigma)$, this becomes an uncoupled Poisson equation, where modes are coupled only to their complex conjugates. This is the functional form of the continuum Lagrangian, which is invariant under translations and therefore conserves momentum. For a more general shape function, $\mathcal{K}$ has off-diagonal elements that couple different modes. This matrix must be inverted once at the beginning of the simulation to compute the fields. It also requires evaluating various convolution integrals between the particle and field shapes, which may prove computationally prohibitive if an analytical expression is not available.

\section{Discretizing the action and the discrete equations of motion}\label{algorithm}

The continuous-time action integral is now approximated by a discrete Lagrangian which is still a function of continuous time:
\begin{equation}
\mathcal{S} \approx \int dt \mathbf{L}_D [\x_j, \vel_j; \tilde{\varphi}_\sigma].
\end{equation}
To find the discrete equations of motion, we must approximate the integral for $\mathcal{S}$ using discrete steps in time, using Riemann sums or higher order approximations. The most direct approach is to use composition methods to make a self-adjoint approximation of the action (see \S 2.4 and \S2.5 of~\cite{marsden_west:01}).

A first order approximation of the action is the Riemann sum
\begin{equation}
\mathbf{S}_D^{(n)} = h \mathbf{L}_D \left [ \x_j^{(n+1)}, \frac{\x_j^{(n+1)} - \x_j^{(n)}}{h}; \tilde{\varphi}_{\sigma}^{(n+1)} \right ]
\end{equation}
for a step size $h$. This is the action for the $n^{th}$ step, and the total action is approximated by $\mathcal{S} \approx \sum_n \mathbf{S}_D^{(n)} + \mathcal{O}(h^2)$. This is a first order approximation. To reach second order accuracy in time, we use a composite scheme of a step with its adjoint. The adjoint of the discrete action is given by
\begin{equation}
\mathbf{S}_D^*(\x^{(n+1)}, \x^{(n)}, h) = -\mathbf{S}_D (\x^{(n)}, \x^{(n+1)}, -h)
\end{equation}
and it is straightforward to show that a self-adjoint action is of an even order of accuracy -- thus by being self-adjoint we are assured at least second order accuracy in time.

By taking the composite action for the $n^{th}$ step
\begin{eqnarray}
\nonumber \mathbf{S}_D = & \underbrace{\sum_i \frac{1}{2} m w_i \left \{  \frac{\left (\mathbf{x}_i^{(n+1)} - \mathbf{x}_i^{(n+\nicefrac{1}{2})}\right )^2}{\nicefrac{h}{2}} + \frac{\left (\mathbf{x}_i^{(n+\nicefrac{1}{2})} - \mathbf{x}_i^{(n)}\right )^2}{\nicefrac{h}{2}} \right \}}_{\textrm{ptcl. move}} \\
\nonumber& - \underbrace{h q\sum_i \sum_\sigma w_i \tilde{\varphi}_{\sigma}^{(n+\nicefrac{1}{2})} \int d\kvec~ \tilde{\Lambda}^* (\kvec) \Psi(\kvec - \kvec_\sigma) e^{i \kvec \cdot \x_i^{(n+\nicefrac{1}{2})}}}_{\textrm{interpolation/deposition}} 
\\
&+ \underbrace{\frac{h}{8 \pi} \sum_{\sigma, \sigma'} \tilde{\varphi}^{(n+\nicefrac{1}{2})}_{\sigma} \mathcal{K}_{\sigma \sigma'} \tilde{\varphi}^{(n+\nicefrac{1}{2})}_{\sigma'}}_{\textrm{Poisson equation}}
\end{eqnarray}
and minimizing the action using the discrete Euler-Lagrange equations (see \S 2.5 of \cite{marsden_west:01}) we get the discrete equations of motion
\numparts
\begin{eqnarray} \label{del_1}
\nonumber m \frac{\x^{(n+1)}_j - 2 \x^{(n+\nicefrac{1}{2})}_j + \x^{(n)}_j}{\nicefrac{h}{2}} = \dots\\
\dots - hei \sum_{\sigma} \int d\kvec ~ \kvec  \tilde{\Lambda}^*(\kvec) \Psi(\kvec - \kvec_\sigma) \tilde{\varphi}_\sigma e^{i \kvec \cdot \x^{(n+\nicefrac{1}{2})}_j} \\ \label{del_2}
\sum_{\sigma'}\mathcal{K}_{\sigma \sigma'} \tilde{\varphi}_{\sigma'} = 8 \pi e \sum_j w_j  \int d\kvec \tilde{\Lambda}^*(\kvec)  e^{i \kvec \cdot \x_j^{(n+\nicefrac{1}{2})}} \Psi(\kvec - \kvec_\sigma)\\
\frac{\x^{(n+\nicefrac{3}{2})} - \x^{(n+1)}}{\nicefrac{h}{2}} = \frac{\x^{(n+1)} - \x^{(n+\nicefrac{1}{2})}}{\nicefrac{h}{2}}
\end{eqnarray}
\endnumparts
where $\mathcal{K}_{\sigma \sigma'}$ is the generalization of the usual $\kvec^2$ in the Poisson equation. The first equation is simply $m \ddot{\x} = q \mathbf{E}$, while the second is the Poisson equation for computing the potential in the middle of the step. The third equation tells us how the boundaries between steps are related. In our case it is the continuity of the ``velocity'', although different discretizations may have different boundary terms.

Note that there is no velocity co\"{o}rdinate -- as discussed by Marsden and West~\cite{marsden_west:01} there is no tangent space in the discrete Lagrangian picture. We may, however, introduce the velocity as a convenient intermediate computational variable. If we define 
\begin{equation}
\vel^{(n)} = \frac{\x^{(n+\nicefrac{1}{2})} - \x^{(n)}}{\nicefrac{h}{2}}
\end{equation}
then the above equations become
\numparts
\begin{eqnarray}
\x^{(n+\nicefrac{1}{2})}_j = \x^{(n)}_j + \frac{h}{2} \vel^{(n)}_j \\
\sum_{\sigma'}\mathcal{K}_{\sigma \sigma'} \tilde{\varphi}_{\sigma'} = 8 \pi e \sum_j w_j  \int d\kvec \tilde{\Lambda}^*(\kvec)  e^{i \kvec \cdot \x_j^{(n+\nicefrac{1}{2})}} \Psi(\kvec - \kvec_\sigma)\\
\vel^{(n+\nicefrac{1}{2})}_j - \vel^{(n)}_j = -\frac{h e}{m} i \sum_{\sigma} \int d\kvec ~ \kvec  \tilde{\Lambda}^*(\kvec) \Psi(\kvec - \kvec_\sigma) \tilde{\varphi}_\sigma e^{i \kvec \cdot \x^{(n+\nicefrac{1}{2})}_j} \\
\x^{(n+1)}_j = \x^{(n+\nicefrac{1}{2})}_j + \frac{h}{2} \vel^{(n+\nicefrac{1}{2})}_j\\
\vel^{(n+1)}_j = \vel^{(n+\nicefrac{1}{2})}_j
\end{eqnarray}
\endnumparts
which is the split-step move and accelerate familiar from most particle-based algorithms. We have avoided using a grid to deposit charge to preserve the translational invariance. This means that every particle has an exponential, $e^{i \kvec \cdot \x}$, that must be calculated.

We have thus far left $\Psi$ almost completely general. To understand conditions on momentum conservation, we derive the total force between two particles. First, let us assume that we have chosen our discrete $\kvec_{\sigma}$ such that there exists a $\kvec_{-\sigma} = - \kvec_{\sigma}$. This is a reasonable choice if we want to make the shape functions $\Lambda$ and $\Psi$ even functions and still return real values for $\varphi(\x)$.

If we place a single particle at position $\x_j$, it will create the potential
\begin{eqnarray}
\nonumber \tilde{\varphi}_{\sigma} &=& 8 \pi e w_j \sum_{\sigma'} \mathcal{K}_{\sigma' \sigma}^{-1} \int d\kvec \tilde{\Lambda}^*(\kvec) e^{i \kvec \cdot \x_j} \Psi(\kvec - \kvec_{\sigma'})\\
 &=& 8 \pi e w_j \sum_{\sigma'} \mathcal{K}_{\sigma' \sigma}^{-1} f_{\sigma'}(\x_j)
\end{eqnarray}
where
\begin{equation}
f_{\sigma}(\x) = \int d \kvec \tilde{\Lambda}^* (\kvec) \Psi(\kvec - \kvec_\sigma) e^{i \kvec \cdot \x_j}
\end{equation}
The force on particle $\ell$ due to particle $j$ is then given by
\begin{equation}
\Delta \mathbf{p}_{j \rightarrow \ell} = 8 \pi h e^2 w_\ell w_j \nabla_{\x_\ell} \sum_{\sigma \sigma'} f_{\sigma'}(\x_j) \mathcal{K}_{\sigma \sigma'}^{-1} f_{\sigma}(\x_\ell)
\end{equation}

If we define
\begin{equation}
F \left (\kvec, \kvec' \right ) = \tilde{\Lambda}^*(\kvec) \tilde{\Lambda}^*(\kvec') i (\kvec + \kvec') e^{i (\kvec \cdot \x_j + \kvec' \cdot \x_\ell)}
\end{equation}
then we can then say that the total change of the momentum for this two-body pair is
\begin{eqnarray}
\nonumber \Delta \mathbf{p}_{j \rightarrow \ell} + \Delta \mathbf{p}_{\ell \rightarrow j} & = A \int_{-\infty}^{\infty} d \kvec d \kvec' ~ F(\kvec, \kvec') \sum_{\sigma, \sigma'} \Psi(\kvec - \kvec_\sigma) \mathcal{K}_{\sigma \sigma'}^{-1} \Psi(\kvec' - \kvec_{\sigma'})\\
\nonumber & = A \int_{-\infty}^{\infty} d \kvec d \kvec' ~ F(\kvec, \kvec') \sum_{\sigma, \sigma'} \Psi(\kvec - \kvec_{-\sigma}) \mathcal{K}_{-\sigma -\sigma'}^{-1} \Psi(\kvec' - \kvec_{-\sigma'}) \\
& = A \int_{-\infty}^{\infty} d \kvec d \kvec' ~ F(\kvec, \kvec') \sum_{\sigma, \sigma'} \Psi(\kvec + \kvec_{\sigma}) \mathcal{K}_{\sigma \sigma'}^{-1} \Psi(\kvec' + \kvec_{\sigma'})
\end{eqnarray}
where we have used to symmetry of $\mathcal{K}_{-\sigma, -\sigma'} = \mathcal{K}_{\sigma, \sigma'}$ and the fact that $\sigma$ and $\sigma'$ are dummy indices. We can then flip the sign on $\kvec$ and $\kvec'$ and exploit the fact that $\Psi$ is an even function to get that
\begin{eqnarray}
\nonumber \Delta \mathbf{p}_{j \rightarrow \ell} + \Delta \mathbf{p}_{\ell \rightarrow j} &= A \int_{-\infty}^{\infty} d \kvec d \kvec' ~ F(-\kvec, -\kvec') \sum_{\sigma, \sigma'} \Psi(\kvec - \kvec_{\sigma}) \mathcal{K}_{\sigma \sigma'}^{-1} \Psi(\kvec' - \kvec_{\sigma'}) \\
&= (\Delta \mathbf{p}_{j \rightarrow \ell} + \Delta \mathbf{p}_{\ell \rightarrow j})^*
\end{eqnarray}
This proves that the total change in momentum is a real number if we choose even functions for our particle and field shapes, and have negative pairs of Fourier modes. However, this will not generally be zero, and we can have some change in the total momentum. This is because the resulting discrete Lagrangian is not, in general, unchanged by translations, so it does not have the associated Noether conserved quantity.

For the special case of a delta function, the Lagrangian is unchanged under a translation, and we can show immediately that momentum is conserved to machine precision. Here, $\mathcal{K}_{\sigma \sigma'} = \kvec_{\sigma} \cdot \kvec_{\sigma'} \delta_{\sigma, -\sigma'}$. There are no self-forces
\begin{equation}
\Delta \mathbf{p}_{j \rightarrow j} = 4 \pi h e^2 w i \sum_{\sigma} \frac{\kvec_{\sigma}}{|\kvec_\sigma|^2} e^{i \kvec_\sigma \cdot \x_j} e^{-i \kvec_{\sigma} \cdot \x_j} = 0
\end{equation}
since every $\kvec_\sigma$ is paired with a $-\kvec_{\sigma} = \kvec_{-\sigma}$, a requirement on the expansion to keep $\varphi(\x)$ real. Because there are no self-forces, we can write the force on particle $\ell$ from particle $j$ as
\begin{equation}
\Delta \mathbf{p}_{\ell \rightarrow j} = 4 \pi h e^2 w_\ell w_j i \sum_\sigma \kvec_{\sigma} e^{i \kvec_{\sigma} \cdot \x_j} e^{-i \kvec_{\sigma} \cdot \x_{\ell}} \frac{1}{|\kvec_\sigma|^2} = - \Delta \mathbf{p}_{j \rightarrow \ell}.
\end{equation}
We must therefore balance our desire for preventing the ringing from using discrete Fourier modes with how much our problem requires exact or nearly-exact momentum conservation. It remains to see whether such systems will grow monotonically in momentum, or whether the total momentum will simply oscillate around the initial total momentum, and how these depend on our shape functions. For now, we will focus on the discrete modes with delta-function shape, as they are the least computationally demanding due to the lack of any numerical integrals in the update sequence.

\section{Numerical Dispersion Relation}\label{numerical_dispersion}

We now derive a linearized dispersion relation for the finite time step spectral algorithm, extending the arguments of \S 9-2 of~\cite{birdsall_langdon:85} for our spectral algorithms. We begin by defining $\x = \mathbf{X} + \delta \x$, where $\mathbf{X}$ is the unperturbed ballistic motion, $\mathbf{X}^{(n)} = \x^{(0)} + \vel^{(0)} n h$ and $|\kvec \cdot \delta \x| \ll 1$ is a small perturbation. 

To include the boundary terms, we must include the update sequence from the end of the last ``accelerate'' to the beginning of the next. For convenience, we will denote these with integer indices, since they are all offset by a half-step. Inserting this approximation into eqn.~(\ref{del_1}), along with our re-indexing, gives
\begin{equation}
\frac{\delta \x_j^{(n+1)} - 2 \delta \x_j^{(n)} + \delta \x_j^{(n-1)}}{h} = - \frac{hei}{m} \sum_{\sigma} \int d\kvec ~ \kvec  \tilde{\Lambda}^*(\kvec) \Psi(\kvec - \kvec_\sigma) \tilde{\varphi}_\sigma e^{i \kvec \cdot \mathbf{X}^{(n)}_j}
\end{equation}
 
 It is interesting to look at a finite width in $\Psi$. This behaves like a finite width in $\vel$ in the particle distribution, creating numerical Landau damping even for a cold plasma, with a damping time $\tau^{-1} \sim \sigma_ \kvec \cdot \vel_0$ for a typical width $\sigma_ \kvec$ to the shape function $\Psi$. We can intuitively understand this as follow: The finite width Fourier shapes represent a distribution of phase velocities. Thus, by introducing these shapes, we create a spread of phase velocities that a given mode couples to the distribution of particles, which is an unphysical but identical configuration to the usual mechanism for physical Landau damping. This may be useful in some scenarios. However, from here on we will consider only the discrete Fourier modes, $\Psi(\kvec - \kvec_\sigma) = \delta(\kvec - \kvec_\sigma)$.
 
  Let us assume that the electrostatic field varies as $\tilde \varphi_\sigma = \tilde{\Phi}_\sigma e^{-i \omega n h}$. The integrand of the right hand side of this equation varies as $e^{i \kvec_\sigma \cdot (\x_j^{(0)} - \vel_j^{(0)} n h) - i \omega n h}$, and so too must $\delta \x$. Under this assumption, we have the dynamics for $\delta \x$ given by
\begin{eqnarray}
\nonumber \frac{\delta \x_j^{(n+1)} - 2 \delta \x_j^{(n)} + \delta \x_j^{(n-1)}}{h} = \\ - \frac{hei}{m} \sum_{\sigma}  \kvec_\sigma  \tilde{\Lambda}^*(\kvec_\sigma) \tilde{\Phi}_\sigma e^{i \kvec_\sigma \cdot (\mathbf{x}^{(0)}_j + \vel_j^{(0)}nh)}e^{-i \omega n h}
\end{eqnarray}
It is convenient to define
\begin{equation}
\delta \x^{(n)} = \sum_\sigma \delta \tilde \x^\sigma e^{-i \omega_d^\sigma n h}
\end{equation}
which in turn gives that
\begin{equation}
\delta \tilde \x_j^\sigma = -\frac{1}{2 (\cos \omega_d^\sigma h - 1)} \frac{h^2 e i}{m}  \kvec_\sigma \tilde \Lambda^*(\kvec_\sigma) \tilde \Phi_\sigma e^{i \kvec_\sigma \x_j^{(0)}}
\end{equation}

It remains to compute $\tilde \Phi_\sigma$. Looking to eqn.~(\ref{del_2}) and Taylor expanding the right hand side gives
\begin{equation}
|\kvec_\sigma|^2 \tilde \Phi^*_{\sigma} = 4 \pi e \sum_j w_j \tilde{\Lambda}^*(\kvec_\sigma) e^{i \kvec_\sigma \cdot (\x_j^{(0)} + \vel_j^{(0)}n h)} e^{i \omega n h} i \kvec_\sigma \cdot \delta \x_j
\end{equation}
where we have assumed that the unperturbed quantity
\begin{equation}
\sum_j w_j \tilde{\Lambda}^*(\kvec_\sigma) e^{i \kvec_\sigma \cdot (\x_j^{(0)} + \vel_j^{(0)}n h)} \approx 0
\end{equation}
which is to say that the unperturbed system has approximate charge neutrality.

Combining the solution for the perturbed orbit with the Poisson equation yields
\begin{eqnarray}
\nonumber |\kvec_\sigma|^2 \tilde{\Phi}_\sigma = & 4 \pi \frac{h^2 e^2}{m} \sum_{j, \sigma'} w_j \tilde{\Lambda}^*(\kvec_\sigma) e^{-i (\kvec_\sigma-\kvec_{\sigma'}) \cdot \x_j^{(0)}} \times \\
 & e^{i (\omega_d^\sigma - \omega_d^{\sigma'})n h} \kvec_\sigma \cdot \kvec_{\sigma'} \tilde \Lambda^*(\kvec_{\sigma'}) \tilde \Phi_{\sigma'} \frac{1}{2 (\cos \omega_d^{\sigma'} h - 1)}
\end{eqnarray}
We note that the sum over $j$ can be approximated as an integral over the initial conditions weighted by the local phase space density
\begin{equation}
\sum_j w_j * \mapsto \int d\x^{(0)} d \vel^{(0)} f(\vel^{(0)}) *
\end{equation}
where we have assumed the initial system has no charge perturbations. Noting the orthogonality relation with Fourier modes, we get that
\begin{equation}
\tilde \Phi_\sigma = \frac{4 \pi e^2 n_0}{m} h^2|\tilde \Lambda (\kvec_\sigma)|^2 \int d\vel f(\vel) \frac{1}{2(\cos \omega_d^\sigma h - 1)} \tilde \Phi_\sigma
\end{equation}
where $n_0$ is the unperturbed initial charge density. This yields the numerical dispersion relation
\begin{equation}
1 = \omega_p^2h^2 |\tilde \Lambda(\kvec_\sigma)|^2 \int d \vel f(\vel) \frac{1}{4 \sin^2 \left (\frac{ (\omega - \kvec_\sigma \cdot \vel) h}{2} \right )}
\end{equation}
for each mode $\kvec_\sigma$. In the limit of $h \rightarrow 0$ and $\tilde \Lambda = 1$ (point particles) this returns the continuum limit. This is as in~\cite{birdsall_langdon:85} but for discrete time steps but continuous space, although the discrete Fourier representation picks specific $\kvec_\sigma$, and with the particle shape functions appearing to modify the dispersion relation. We can carry out additional analysis on this dispersion relation.

Immediately we see that this algorithm modifies the plasma frequency for each mode to 
\begin{equation}
\omega_p^\sigma = \omega_p |\tilde{\Lambda}(\kvec_\sigma)|.
\end{equation}
This requires that $|\tilde \Lambda(\kvec_\sigma)|$ stay close to unity until the Debye wavenumber, which is equivalent to requiring that the particle shape function $\Lambda(\x)$ be narrower than the Debye length.

Assuming a cold coasting plasma, $f(\vel) = \delta(\vel - \vel_0)$, the dispersion relation becomes
\begin{equation}
4 \sin^2 \left ( \frac{(\omega - \kvec_\sigma \cdot \vel_0)h}{2} \right ) = \omega_p^2 h^2 |\tilde \Lambda(\kvec_\sigma)|^2
\end{equation}
which in turn requires that
\begin{equation}
\frac{\omega_p h}{2} |\tilde \Lambda(\kvec_\sigma)| < 1
\end{equation}
which is the modified $\omega_p-\Delta t$ condition for our electrostatic algorithm. This also gives a guideline for accurate spectral fidelity for the linear plasma dispersion relation (must be much smaller than $1$), and the restriction is most stringent for long wavelength modes, assuming $|\tilde \Lambda(\kvec_\sigma)|$ has a decreasing envelope with $|\kvec_\sigma|$.

We can furthermore use Nyquist diagrams to evaluate the numerical Penrose criterion, since the dispersion relation has poles at $\omega - \kvec_\sigma \cdot \vel = n \pi$. However, only the $n=0$ pole will contribute, as the residues of all the others are zero. Since dynamics is only parallel to $\kvec_\sigma$, let us limit ourselves to a one-dimensional dispersion relation parallel to $\kvec_\sigma$ for simplicity:
\begin{equation}
1 = \omega_p^2h^2 |\tilde \Lambda(\kvec_\sigma)|^2 \int d v \frac{f(v)}{(\omega - k_\sigma v)^2} \frac{1}{4 \left ( \frac{\sin \left (\frac{ (\omega - k_\sigma v) h}{2} \right )}{\omega - k_\sigma v} \right )^2}.
\end{equation}
This integral is given by the Cauchy principle value integral
\begin{equation}
1 = \omega_p^2 h^2 |\tilde{\Lambda}(\kvec_\sigma)|^2 \left (\mathcal{P} \int dv \dots + \pi i \frac{f'(\nicefrac{\omega}{k_\sigma})}{h^2 k_\sigma^2} \right ).
\end{equation}
The imaginary part is of interest here. Because it has the identical functional form to the continuum limit with a modified plasma frequency, we can conclude that there are no additional plasma instabilities introduced by the algorithm so long as the $\omega_p-\Delta t$ condition is satisfied.

That instability arises in the $\mathcal{P}$ part of the integral, which must be able to equal $1$ for real values of $\omega$. The sinusoid in the denominator may prevent that due to the repeated pole structure, although the details will depend on $f(\vel)$. If we assume a cold but not infinitely cold plasma, and a gaussian distribution $f(v) = \frac{\exp(-\nicefrac{v^2}{2 \sigma_v^2})}{\sqrt{2 \pi \sigma_v^2}}$, with $\sigma_v \ll \nicefrac{1}{h k_\sigma}$, then we can approximate the Cauchy principal value asymptotically as
\begin{equation}
\mathcal{P} \int dv \dots \approx \frac{1}{4 \sin^2 \frac{\omega h}{2}} + \mathcal{O}(h k_\sigma \sigma_v)
\end{equation}
which means thermal effects offer a small modification of to the cold $\omega_p-\Delta t$ condition. It also offers an additional guideline for accurate modeling, which is that for an initially thermal distribution $|h k_\sigma \sigma_v| \ll 1$ for the largest values of $k_\sigma$ or else we could introduce a numerical plasma instability with a sufficiently large time step.

\section{Numerical Example -- Two-Body Problem}\label{twobody}

As a first test of the angular and linear momentum conserving properties of this algorithm, we consider the two dimensional Kepler problem. In two dimensions the particles are line charges, and the electrostatic potential is given by $V(r) \sim \ln (\nicefrac{r}{r_0})$, so this differs from the standard $\nicefrac{1}{r}$ Kepler problem. Our test problem is a single electron $(m_e, e)$ and a ``double-positron'' $(2 m_e, -2 e)$. There is some initial total linear momentum. We use tent functions for the particle shapes and discrete Fourier modes, with $\Psi(\kvec) = \delta(\kvec)$, with a time step of $\unit[5. \times 10^{-9}]{sec}$ for $2 \times 10^6$ time steps. For the given initial conditions, this is approximately 9-10 steps per period, which reasonably resolves the dynamics of the problem.

\begin{figure} 
\begin{center}
\includegraphics[scale=0.5]{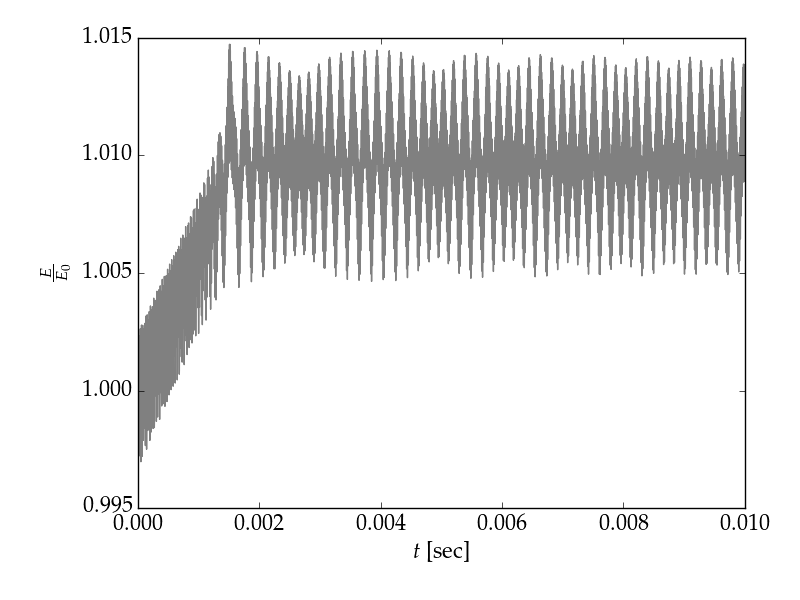}
\end{center}
\caption{Total energy relative to the initial energy of the two-body system.}
\label{energy}
\end{figure}

\begin{figure} 
\begin{center}
\includegraphics[scale=0.5]{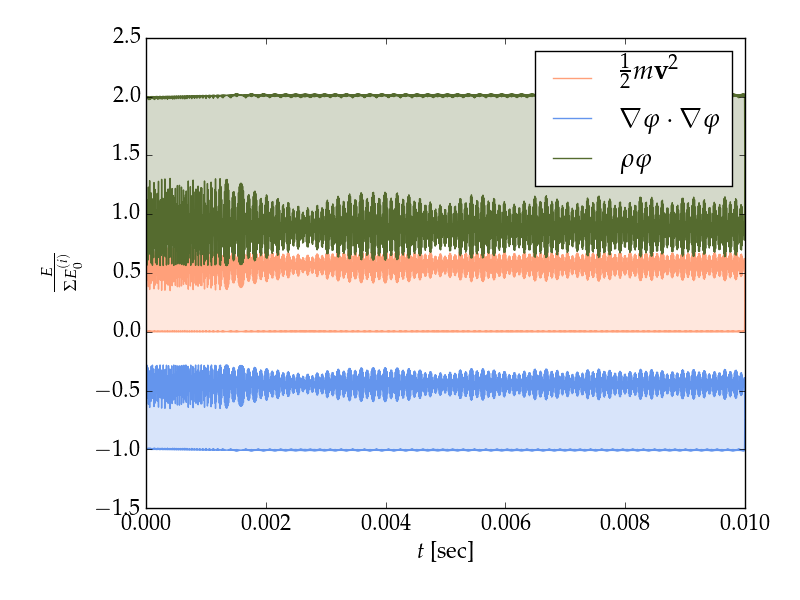}
\end{center}
\caption{The individual contributions of the kinetic, capacitive, and field energies to the total energy of the Kepler problem.}\label{energy_envelopes}
\end{figure}

As is well-known, if one calculates the energy of a conservative Hamiltonian system from simulations using a symplectic integrator, that energy is bounded for exponential time. We see this behavior in fig. (\ref{energy}). After some initial transient increase in the energy, it is conserved at the \unit[1]{\%} level over an extended time. We can also look at the individual energy contributions -- kinetic, capacitive, and field energies -- to see that each is very accurately bounded, so that there is no drift in the individual components that happens to cancel. The larger scale oscillations are a result of computing the energy every 250 time steps, i.e. once every 25 periods or so. This creates much of the rippling we see in fig. (\ref{energy_envelopes}).

Fig. (\ref{energy_envelopes}) demonstrates that the energy conservation does not arise from spurious growth in kinetic energy and decay in potential energy, or \emph{vice versa}. As one would expect, the individual contributions oscillate and are well-behaved. The energy conservation then does not arise from any spurious trends in the individual contributions which happen to cancel out on the whole.

Furthermore, the robustness of the behavior is independent of the timing of the co\"ordinates during the time step when the energy is computed. This means that the energy evaluated at the end of a full time step has the same level of precision (although will differ numerically) as the energy evaluated if a half-drift is used to put $\x$ and $\vel$ at the same time, and then the fields are computed using $\x$. This is another property familiar from single particle symplectic integrators.

\begin{figure} 
\begin{center}
\includegraphics[scale=0.5]{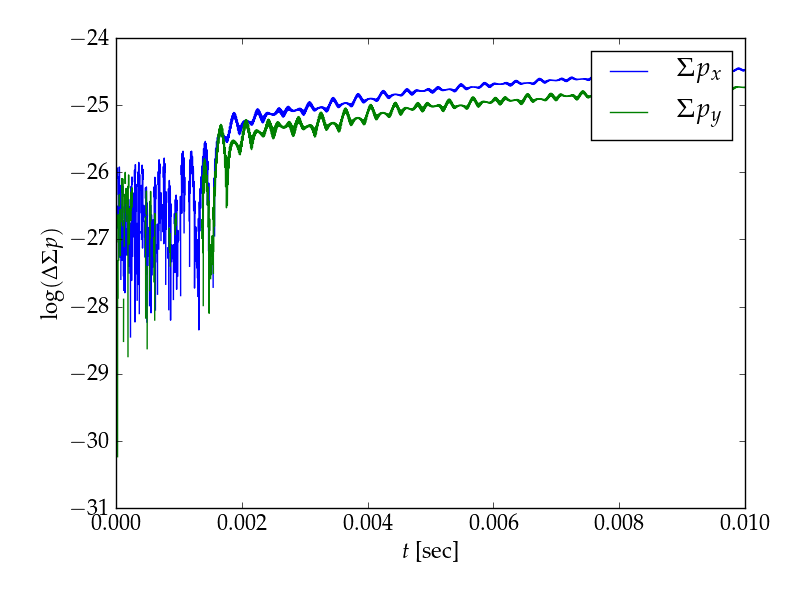}
\end{center}
\caption{Total $p_x$ and $p_y$ for the system, in a $\log$ scale.}
\label{momentum}
\end{figure}

A major feature of this discrete Fourier spectral algorithm is that it is translationally invariant, therefore it should conserve the total momentum. In fig. (\ref{momentum}) we see that the initial total linear momentum is conserved to machine precision. There is an initial nonzero total momentum, and for the first five hundred thousand time steps the momentum change is machine precision zero. The upward trend after this point is likely due to roundoff error.

\begin{figure} 
\begin{center}
\includegraphics[scale=0.5]{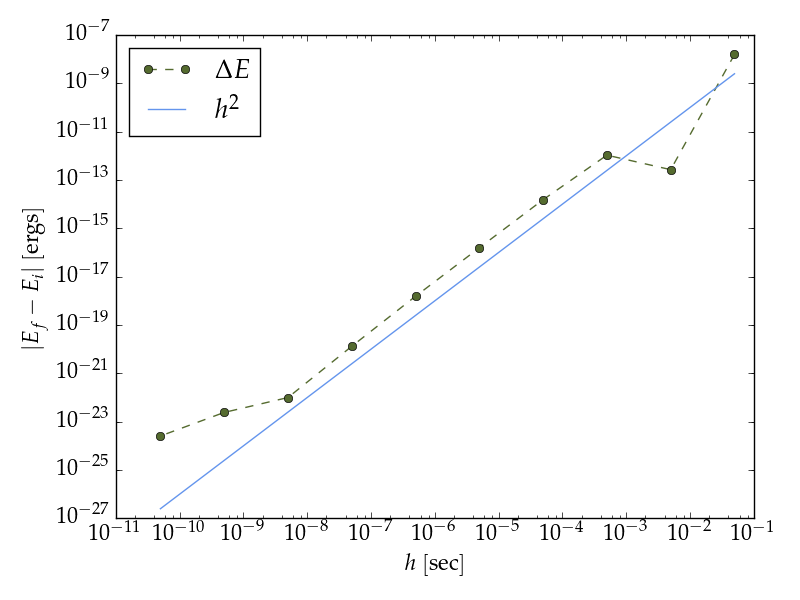}
\end{center}
\caption{Energy error over a single time step versus $h$.}
\label{energyscaling}
\end{figure}

The algorithm also purports to be second order accurate, meaning that the correction to the continuous action integral is given by $\mathbf{S}_D \approx \mathcal{S} + \mathcal{O}(h^3)$. Based on scaling, where $\mathcal{S} \sim \int E dt$, this would imply that the correction to the energy would be $E_D \approx E + \mathcal{O}(h^2)$. For the two-body problem, we can clearly see this in fig.~(\ref{energyscaling}).

As can be seen, the energy deviation scales with $h^2$ for this particular problem from around \unit[$10^{-3}$]{sec.} where the failure to resolve the physics becomes catastrophic\footnote{At this point, the relative energy error is $\sim 10^3-10^4$.}, down to around \unit[$10^{-8}$]{sec} where machine precision issues come into play. This clearly demonstrates the second order accuracy of the spectral algorithm. It should be noted that this algorithm was implemented using NumPy (v. 1.9.2) in Python (v. 2.7.10) on a 64-bit machine -- handling of machine precision will vary across architectures and programming languages.

We have thus shown that we can accurately simulate the two-body Kepler problem for a hundred thousand periods of the system with no change in the total momentum and with very well-behaved energy which does not spuriously grow. Such a problem is very difficult to do in conventional gridded PIC, where the total energy grows and the total momentum is changing, which breaks the careful conservation that allows for a closed orbit two-body problem. We see that this multisymplectic integrator preserves a number of key geometric properties of the system -- the angular and linear momentum are conserved exactly, with good behavior for the total energy due to the nature of symplectic algorithms.

\section{Numerical Example -- Thermal Plasma}~\label{thermal_plasma}

Having demonstrated the energy and momentum conserving properties of this algorithm for a two-body problem, we now consider a thermal plasma. Again, we consider periodic boundaries to keep the system closed and make the total energy a quantity that is conserved in the physical system. In conventional PIC, such a system would show spurious energy growth over many plasma periods.

We show simulation results run for $\sim 10^6$ plasma oscillations. We consider a system with the parameters in fig. (\ref{simparams}), which is a very low resolution simulation. We are resolving Debye length physics here -- in practice the spectral method introduces an ultraviolet cutoff on the resolved physics. Any length shorter than $(k_{\textrm{max}})^{-1}$ is removed entirely, as can be verified by failing to resolve the Debye length.

\begin{figure}
\begin{center}
\begin{tabular}{|c | c|}
\hline
Quantity & Value [units] \\
\hline
electron density & $\unit[10^{17}]{cm^{-3}}$\\
$\lambda_D$ & $\unit[3.96\times 10^{-11}]{cm}$ \\
$\omega_p$ & $\unit[8.14 \times 10^{17}]{sec^{-1}}$\\
\hline
$L_{\textrm{domain}}$ & $4 \times \lambda_D$\\
$n_{\textrm{macro}}$ & 16\\
macro. weight & $9.9 \times 10^5$\\
macro. width & $0.1 \lambda_D$ \\
$\delta k$ & $\nicefrac{\pi}{2} (L_{\textrm{domain}})^{-1}$\\
$n_{\textrm{modes}}$ & 80\\
$h^{-1}$ & $5 \times \nicefrac{\omega_p}{2\pi}$\\
\hline
\end{tabular}
\caption{First simulation parameters.}\label{simparams}
\end{center}
\end{figure}

\begin{figure} 
\begin{center}
\includegraphics[scale=0.5]{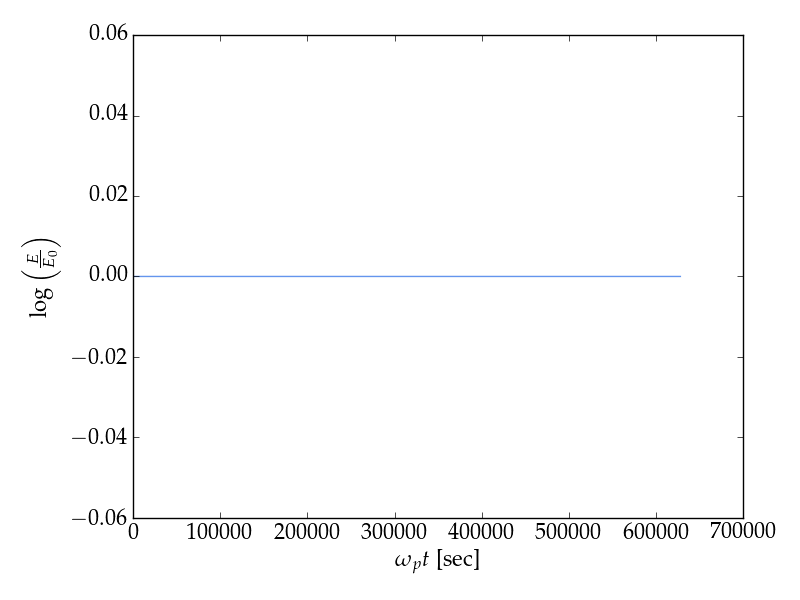}
\end{center}
\caption{Energy change over time for thermal plasma simulation.}
\label{thermal1energy}
\end{figure}

\begin{figure} 
\begin{center}
\includegraphics[scale=0.5]{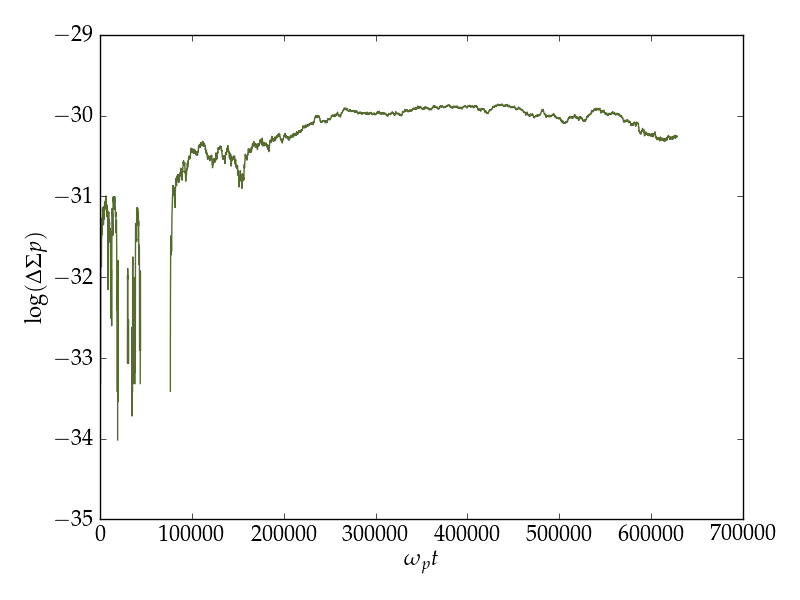}
\end{center}
\caption{Momentum change over time for thermal plasma simulation.}
\label{thermal1momentum}
\end{figure}

In this simulation, there is no discernible change in the total energy of the system (fig. (\ref{thermal1energy})). This is to be expected, as the fields should be zero in the physical system, and we would expect all the motion to be ballistic. In two or three dimensions, we do not anticipate this level of energy conservation, which likely arises from a lack of degrees of freedom to cause the energy oscillations observed in the two-dimensional two-body problem. We also see the same momentum conservation, fig. (\ref{thermal1momentum}), which oscillates near machine precision throughout the simulation.

In conventional particle-in-cell simulations, even in one dimension, it is known that the plasma will heat up~\cite{ueda_etal:94, mardahl_thesis}. As noted in \cite{mardahl_thesis}, various techniques in conventional algorithms -- such as higher order particle shapes or increasing the grid resolution of the Debye length -- can reduce the numerical heating up to a point. This algorithm removes the heating entirely, as well as dramatically reducing the statistical fluctuations intrinsic in conventional particle-in-cell.

The reduction in statistical fluctuations can be understood by looking at how the variational method would replicate a conventional algorithm. Conventional algorithms have no intrinsic convolution integrals, so we can understand the ``particle shape'' as either a particle shape with delta function fields, or a ``field shape'' with delta function particles. Both options are extremely noisy. By introducing the convolution integrals and using a Fourier basis, we smooth the particle shapes while simultaneously removing high frequency content without aliasing. This has been demonstrated in finite-difference versions of variational algorithms in, e.g., \cite{shadwick:14a}, where a one dimensional electromagnetic finite differencing shows extremely smooth charge distributions for the applied problem. This comes from using a combination of field and particle shape functions, instead of just one or the other.

\section{Implementation and Optimization}~\label{implementation}

This algorithm has a number of properties that differ from conventional particle-in-cell methods that requires some consideration. This includes the discrete mode properties, parallel implementation, the global nature of the field solve, and evaluation of the deposition/interpolation terms. Also, the spectral nature of the algorithm requires some care to minimize the computational costs.

The first is the span of $k$-space. Implemented with periodic boundaries, the longest wavelength is $L$, the simulation domain length. Furthermore, we must resolve adequately the particle width, $\ell$. Thus, our $k$-space modes must span $k \in [2 \pi/L, \sim 2\pi/\ell]$. Furthermore, because the system has periodic boundaries, we must have as our increment in $k$-space $\Delta k = 2 \pi/L$ to maintain periodicity. Thus, we must use $\mathcal{O}(L/\ell)^D$ Fourier modes. This is comparable to the number of cells required for a traditional gridded method.

Because the algorithm is truly spectral (there is no grid in real space) conventional domain decomposition is irrelevant. This simplifies parallel implementations. Each process must have all the field data, and therefore the field solve can be reduced to a single \texttt{MPI\_AllReduce} call on the $\tilde{\varphi}_\sigma$ after the deposition phase. All other operations can be performed locally. This also makes load balancing trivial -- every process will receive an equal number of macro-particles and store all the field information. However, this could present a challenge for very large simulations with substantial amounts of field data. Because the simulation requires no overhead for messaging particle data, and is perfectly load balanced, such spectral algorithms may be ideal for simulations of devices such as magnetrons which have fields everywhere but particles which are isolated to a small region of space throughout much of the simulation.

Computing the deposition and interpolation can be computationally intensive due to the exponential evaluations. This can be optimized in two ways for the discrete Fourier treatment (Fourier shape functions cannot avoid this problem). The first is by recognizing that we are always taking the real part of the fields, so we only have to keep the $\vec{k}_\sigma$ with all components greater than or equal to zero. This is a factor of $2^D$ savings in the amount of information that must be stored. We can also reduce the number of exponentials per particle from $N_{modes}$ to $D$ if we use regularly spaced Fourier modes spaced by $\nicefrac{2 \pi}{L}$ by tracking $\nicefrac{2 \pi x}{L}$. Then we must evaluate the exponential once, and each coefficient is simply a positive integer power of that exponential. Compared to the the na\"{i}ve algorithm, this can represent a substantial savings in computation and memory footprint.

\section{Conclusion}

We have derived an electrostatic spectral macroparticle algorithm from a variational approach. The resulting algorithm closely resembles conventional leapfrog algorithms, although we have bypassed using a grid -- hence the choice to refer to it as a ``spectral macroparticle algorithm'' instead of a ``particle in cell algorithm''. A broad class of algorithms are available with finite widths in $k$-space for the field modes. However, our analysis has shown that discrete Fourier modes (rather than modes with finite width) are exactly momentum conserving, and furthermore observed that finite width modes introduce a numerical Landau damping term to the dispersion relation. For these reasons, we consider the discrete Fourier algorithm as the ideal implementation. This algorithm has demonstrated remarkable energy conservation properties, as one would expect from symplectic algorithms.

We have demonstrated the discrete Fourier algorithm prevents all spatial discretization instabilities, and is only prone to the $\omega_p-\Delta t$ instability that arises from time discretization. This is therefore a remedy for the finite grid instability. We have also seen these algorithms demonstrated in two two-dimensional simulations: a two-body problem and a thermal plasma, both with periodic boundaries. These are, to the author's best knowledge, the first computational examples of two-dimensional variational algorithms. They demonstrate proper second-order energy error, as one would expect.

Because we considered Cartesian electrostatics, we were not presented with a number of issues important for electromagnetic algorithms, particularly gauge invariance and the associated charge conservation, as well as difficulties in non-Cartesian co\"{o}rdinate systems such as the pole at the origin of spherical or cylindrical co\"{o}rdinates. This represents future work.

Variational algorithms have a number of advantages over conventional plasma algorithms. The primary one is that the equations of motion come from a discrete Euler-Lagrange equation. Because of this, all approximations (spectral decomposition, co\"{o}rdinate transformations for particles and fields, time discretization) are made at the level of the action -- the discrete Euler-Lagrange equations then construct the algorithm from these approximations. This gives more freedom to the algorithm developer, as the approximations are guaranteed to be self-consistent.

\section{Acknowledgements}

The author would like to thank B. Shadwick (UNL) for fruitful discussions on variational algorithms. The author would also like to thank D. Bruhwiler and N. Cook (RadiaSoft) and J.-L. Vay and R. Lehe (LBL) for invaluable discussions concerning the content of this manuscript.

This work was sponsored by the Air Force Office of Scientific Research, Young Investigator Program, under contract no. FA9550-15-C-0031. Distribution Statement A. Approved for public release; distribution is unlimited.

\bibliography{espicbib}

\end{document}